\documentclass[10pt]{iopart}

\usepackage{graphicx}
\usepackage{iopams}

\begin{document}

\title{Influence of spin and charge fluctuations on spectra of the two-dimensional Hubbard model}

\author{A Sherman}

\address{Institute of Physics, University of Tartu, W. Ostwaldi Str 1, 50411 Tartu, Estonia}

\ead{alekseis@ut.ee}

\begin{abstract}
The influence of spin and charge fluctuations on spectra of the two-dimensional fermionic Hubbard model is considered using the strong coupling diagram technique. Infinite sequences of diagrams containing ladder inserts, which describe the interaction of electrons with these fluctuations, are summed, and obtained equations are self-consistently solved for the ranges of Hubbard repulsions $2t\leq U\leq 10t$, temperatures $0.2t\lesssim T\lesssim t$ and electron concentrations $0.7\lesssim\bar{n}\leq1$ with $t$ the intersite hopping constant.  For all considered $U$ the system exhibits a transition to the long-range antiferromagnetic order at $T_{\rm AF}\approx 0.2t$. At the same time no indication of charge ordering is observed. Obtained solutions agree satisfactorily with results of other approaches and obey moments sum rules. In the considered region of the $U$-$T$ plane, the curve separating metallic solutions passes from $U\approx6t$ at the highest temperatures to $U=2t$ at $T\approx T_{\rm AF}$ for half-filling. If only short-range fluctuations are allowed for the remaining part of this region is occupied by insulating solutions. Taking into account long-range fluctuations leads to strengthening of maxima tails, which transform a part of insulating solutions into bad-metal states. For low $T$, obtained results allow us to trace the gradual transition from the regime of strong correlations with the pronounced four-band structure and well-defined Mott gap for $U\gtrsim6t$ to the Slater regime of weak correlations with the spectral intensity having a dip along the boundary of the magnetic Brillouin zone due to an antiferromagnetic ordering for $U\lesssim3t$. For $T\approx T_{\rm AF}$ and $U\gtrsim 7t$ doping leads to the occurrence of a pseudogap near the Fermi level, which is a consequence of the splitting out of a narrow band from a Hubbard subband. Obtained spectra feature waterfalls and Fermi arcs, which are similar to those observed in hole-doped cuprates.
\end{abstract}

\vspace{2pc}


\maketitle

\ioptwocol

\section{Introduction}
The influence of charge and spin fluctuations on spectra of the fermionic Hubbard model has been attracting considerable attention due to the intimate relation of this problem to the momentum dependence of the electron self-energy and possible orderings of carriers. Short-range fluctuations were considered using Monte-Carlo simulations \cite{Bulut,Haas,Preuss,Grober,Scalapino}, cellular dynamic mean-field theory (CDMFT) \cite{Maier,Kyung,Park,Sato}, cluster perturbation theory (CPT) \cite{Senechal00,Senechal04,Kohno}, dynamical cluster approximation (DCA) \cite{Maier,Huscroft,Moukouri,Merino14}, variation cluster approximation (VCA) \cite{Potthoff,Balzer,Arrigoni,Faye} and strong coupling diagram technique (SCDT) \cite{Sherman17}. More distant fluctuation were taken into account using the dynamic vertex approximation (D$\Gamma$A) \cite{Toschi,Schafer,Rohringer} and dual fermion approach (DF) \cite{Rohringer,Rubtsov08,Hafermann,Rubtsov}. In the spirit of the dynamic mean-field approximation (DMFT) \cite{Georges} the two latter methods use vertices and Green's functions of the Anderson impurity model \cite{Hewson} for calculating infinite sums of ladder diagrams defining spin and charge susceptibilities. The foregoing works were aimed at the description of antiferromagnetic fluctuations, a pseudogap near the Fermi level and refined boundaries of the Mott metal-insulator transition.

In this work, the SCDT \cite{Vladimir,Metzner,Craco,Pairault,Sherman06} is used for investigating the influence of spin and charge fluctuations of all ranges on spectra of the two-di\-men\-si\-o\-nal (2D) repulsive Hubbard model. The approach is based on a regular series expansion of Green's functions in powers of hopping constants, and all calculations are performed for the actual space dimensionality. In previous works \cite{Vladimir,Pairault,Sherman06} it was shown that already two lowest-order diagrams in the expansion of the irreducible part are enough for describing the Mott metal-insulator transition. Moreover, this approximation was demonstrated \cite{Sherman15} to represent spectral functions in a reasonable agreement with Monte Carlo results \cite{Haas,Preuss,Grober} for moderate Hubbard repulsions $U$, temperatures $T$ and different electron concentrations. In particular, at half-filling and low $T$ the intensity in the calculated spectral functions is suppressed near frequencies $\omega=\pm U/2$. In insulating states, together with the Mott gap these two pseudogaps impart a four-band shape to the spectra. Analogous pseudogaps and shapes were observed in spectra in Monte Carlo simulations \cite{Preuss,Grober}. As follows from the obtained equations, the pseudogaps are connected with multiple reabsorptions of carriers with the creation of doubly occupied sites \cite{Sherman15,Sherman16}.

The same idea of a series expansion in powers of hopping constants forms the basis of the diagram technique for Hubbard operators \cite{Zaitsev,Izyumov88,Izyumov90,Ovchinnikov}. This diagram technique is less convenient than the SCDT -- its rules are rather intricate, they and the graphic representation of the expansion vary depending on the choice of the operator precedence. The SCDT is much more compact. For example, the first order for the one-electron Green's function is described by two diagrams in SCDT, while the same processes are depicted by nine diagrams in the technique for Hubbard operators, and the difference in the number of diagrams rapidly grows with increasing order. This makes the calculation of high-order terms a much more elaborate procedure in the technique for Hubbard operators in comparison with SCDT.

Comparing with other approaches for the Hubbard model the following advantages of the SCDT can be noticed: The numerical algorithm is simple and clear; it does not need in time-consuming calculations using powerful computation technique. The convergence of the iteration procedure is rather fast except for the nearest vicinity of the antiferromagnetic transition. In contrast to cluster methods such as CPT and VCA the used approach does not reduce the original translational symmetry of the problem. The CPT is a version of the SCDT, in which cumulants of small clusters are used instead of site cumulants, and an expansion is performed in powers of intercluster hopping constants \cite{Senechal00,Senechal04}. Due to the complexity of calculating second-order cluster cumulants the CPT uses only the first-order cumulants and chain diagrams constructed from them. As a consequence these calculations are not self-consistent, while the SCDT in this work is a self-consistent approach. In contrast to CPT, VCA, CDMFT and DCA the SCDT allows one to consider charge and spin fluctuations of all ranges. As opposed to the D$\Gamma$A and DF the method does not use results of infinite-dimensional Hubbard or Anderson impurity models in consideration of the 2D model. The DMFT grossly overestimates the critical repulsion of the Mott metal-insulator transition, and in the DF a special procedure is used to obtain the correct value \cite{Hafermann}. Besides, one of the distinctive properties of the 2D Hubbard model is the short-range (for $T>0$) antiferromagnetic ordering. In the DMFT this ordering is introduced artificially. In the D$\Gamma$A and DF the ordering is related to the long-range spin fluctuations. However, the calculation uses Green's functions and vertices from DMFT, and it remains unclear which of them -- from antiferromagnetic or paramagnetic solution -- should be applied for a given temperature. There is no such an uncertainty in the SCDT. Besides, it is an analytic approach, which simplifies the interpretation of obtained peculiarities. Clearly the approach has a number of drawbacks. The series expansion in powers of the kinetic energy implies strong electron correlations. Therefore, the SCDT can be expected to work worse for small $U$. Indeed, in Sec.~3.2 we shall see that the fulfilment of sum rules is impaired in this case. However, taking into account charge fluctuations allows one to minimize this difficulty, as it is demonstrated in Fig.~\ref{Fig11}. Another shortcoming of the SCDT in its present form is a divergence of an irreducible vertex for vanishing $T$. This leads to the finite temperature of the transition to the long-range antiferromagnetic order $T_{\rm AF}\approx 0.2t$. This temperature depends only weakly on $U$. Its finite value is in contradiction with the Mermin-Wagner theorem \cite{Mermin}. We shall discuss below how this defect can be remedied. It is worth noting that in the D$\Gamma$A and DF approximations values of $T_{\rm AF}$ are close to zero.

In SCDT, an interaction of electrons with charge and spin fluctuations is described by diagrams containing ladder inserts. Sums of analogous ladder diagrams define charge and spin susceptibilities. In this work, infinite sums of diagrams with ladder inserts are included into the irreducible part, and obtained equations for the one-particle Green's function are self-consistently solved for the ranges of Hubbard repulsions $2t\leq U\leq 10t$, temperatures $0.2t\lesssim T\lesssim t$ and electron concentrations $0.7\lesssim\bar{n}\leq1$. In this work, the longitudinally irreducible four-leg $ph$ vertex is approximated with its lowest-order term -- the second-order cumulant of electron operators. In spite of the complexity of this quantity, it is possible to reduce summations of the infinite diagram series constructed from these cumulants to the solution of small systems of linear equations. These infinite series of diagrams allow us to take into account interactions of electrons with spin and charge fluctuations of all ranges. As mentioned above, these interactions lead to the transition to long-range antiferromagnetic order.

For some sets of parameters obtained densities of states (DOS) and spectral functions are compared with available results of Monte Carlo simulations \cite{Rubtsov}, exact diagonalization \cite{Dagotto}, DF, DMFT \cite{Rubtsov} and DCA \cite{Huscroft} calculations. The agreement is quite satisfactory. Besides, spectral functions are shown to obey the moments sum rules \cite{Kalashnikov,White,Vilk} with good accuracy. In our calculations, spin and charge fluctuations are taken into account on the same footing. However, in spite of the fact that the approximation describes the ordering in the spin subsystem, no indication of char\-ge ordering is observed in the model. At half-filling, in the considered region of the $U$-$T$ plane a curve separating metallic solutions passes from $U\approx 6t$ at the highest temperatures to $U=2t$ at $T\approx T_{\rm AF}$ (see Fig.~\ref{Fig9}). The value $U\approx 6t$ is close to that obtained for the separation curve in the VCA \cite{Balzer}, CDMFT \cite{Park} and second-order DF \cite{Rohringer}. For moderate $U$ the curve layout resembles that derived in the two-particle self-consistent theory \cite{Vilk}, D$\Gamma$A \cite{Schafer} and ladder DF \cite{Rohringer}. However, if in the two last works, based on the DMFT vertices and Green's functions, the curve location was related to the long-range fluctuations, in our approach it is mainly connected with the short-range fluctuations, as it follows from comparison with results of Ref.~\cite{Sherman17}, in which only these latter fluctuations were allowed for. In this approximation only three types of solutions -- metallic, insulating and critical on the separation curve -- were found, as in Ref.~\cite{Moukouri} using 64-site clusters in the DCA. As will be seen below, the inclusion of long-range fluctuations strengthen tails of spectral maxima, which leads to the appearance of a finite DOS in Mott gaps in a part of formerly insulating solutions, transforming them into bad-metal states. Insulating solutions are retained for $U\gtrsim6t$ and low temperatures. For such $T$, obtained results allow us to trace the gradual transition from the regime of strong electron correlations with the pronounced four-band structure and well-defined Mott gap for $U\gtrsim 6t$ to the Slater regime of weak correlations with a dip in spectra along the boundary of the magnetic Brillouin zone, arising due to the antiferromagnetic ordering, for $U\lesssim3t$. The pseudogaps of the four-band structure are related to regions of a steep dispersion in the electron spectrum, and in the doped case these peculiarities are similar to high-energy anomalies or waterfalls observed in photoemission of several families of cuprates \cite{Ronning,Graf,Valla}. For $T\approx T_{\rm AF}$ and $U\gtrsim7t$ one more pseudogap appears near the Fermi level in the DOS. The pseudogap occurrence is connected with the splitting out of a narrow band from a Hubbard subband. This mechanism resembles the situation in the 2D $t$-$J$ model, in which the spin-polaron band split out of the Hubbard subband is the origin of the pseudogap \cite{Sherman97}. Calculated Fermi surfaces are similar to those observed in photoemission of hole-doped cuprates \cite{Damascelli,Yoshida} -- a non-closed Fermi surface constructed from arcs near nodal points in the underdoped case and a rhombus-shaped closed Fermi surface for the optimal doping.

\section{Main formulae}
The Hamiltonian of the 2D fermionic Hubbard model \cite{Hubbard} reads
\begin{equation}\label{Hamiltonian}
H=\sum_{\bf ll'\sigma}t_{\bf ll'}a^\dagger_{\bf l'\sigma}a_{\bf l\sigma}
+\frac{U}{2}\sum_{\bf l\sigma}n_{\bf l\sigma}n_{\bf l,-\sigma},
\end{equation}
where 2D vectors ${\bf l}$ and ${\bf l'}$ label sites of a square plane lattice, $\sigma=\pm 1$ is the spin projection, $a^\dagger_{\bf l\sigma}$ and $a_{\bf l\sigma}$ are electron creation and annihilation operators, $t_{\bf ll'}$ is the hopping constant and $n_{\bf l\sigma}=a^\dagger_{\bf l\sigma}a_{\bf l\sigma}$. In this work only the nearest neighbor hopping constant $t$ is supposed to be nonzero.

\begin{figure}[t]
\centerline{\resizebox{0.95\columnwidth}{!}{\includegraphics{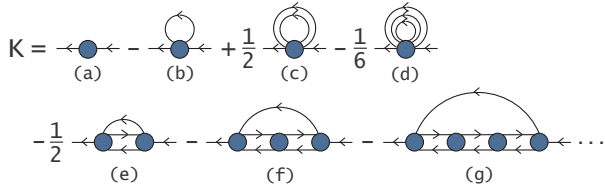}}}
\caption{Diagrams of several lowest orders in $K({\bf k},j)$.} \label{Fig1}
\end{figure}

We shall consider the electron Green's function
\begin{equation}\label{Green}
G({\bf l',\tau';l,\tau})=\langle{\cal T}\bar{a}_{\bf l'\sigma}(\tau')
a_{\bf l\sigma}(\tau)\rangle,
\end{equation}
where the statistical averaging denoted by the angular brackets and time dependencies $$\bar{a}_{\bf l\sigma}(\tau)=\exp{({\cal H}\tau)}a^\dagger_{\bf l\sigma}\exp{(-{\cal H}\tau)}$$
are determined by the operator ${\cal H}=H-\mu\sum_{\bf l\sigma}n_{\bf l\sigma}$ with the chemical potential $\mu$. The time-ordering operator ${\cal T}$ arranges operators from right to left in ascending order of times $\tau$. In this work, for calculating this function the SCDT is used \cite{Vladimir,Metzner,Craco,Pairault,Sherman06} (a concise description of the approach can be found in Refs.~\cite{Sherman17,Sherman16}). In this approach, Green's function is represented by the series expansion in powers of $t_{\bf ll'}$, each term of which is a product of the hopping constants and on-site cumulants of creation and annihilation operators. These terms can be visualized as a sequence of directed lines corresponding to the hopping constants $t_{\bf ll'}$, which connect circles picturing cumulants of different orders. All these terms can be summed in the following expression for the Fourier transform of Green's function (\ref{Green}):
\begin{equation}\label{Larkin}
G({\bf k},j)=\Big\{\big[K({\bf k},j)\big]^{-1}-t_{\bf k}\Big\}^{-1},
\end{equation}
where ${\bf k}$ is the 2D wave vector, $j$ is an integer defining the Matsubara frequency $\omega_j=(2j-1)\pi T$, $t_{\bf k}$ is the Fourier transform of $t_{\bf ll'}$ and $K({\bf k},j)$ is the irreducible part -- the sum of all two-leg irreducible diagrams, which cannot be divided into two disconnected parts by cutting a hopping line. Several lowest order terms of the expansion for $K({\bf k},j)$ are shown in Fig.~\ref{Fig1}.
The linked-cluster theorem is valid and partial summations are allowed in the SCDT. Thanks to this possibility, bare internal lines $t_{\bf k}$ in Fig.~\ref{Fig1} can be transformed into dressed ones,
\begin{equation}\label{hopping}
\theta({\bf k},j)=t_{\bf k}+t^2_{\bf k}G({\bf k},j).
\end{equation}

In this figure, diagrams (a) and (b) contain on-site cumulants of the first and second orders
\begin{eqnarray*}
&&C_1(\tau',\tau)=\big\langle{\cal T}\bar{a}_{{\bf l}\sigma}(\tau')a_{{\bf l}\sigma}(\tau)\big\rangle_0,\\
&&C_2(\tau_1,\sigma_1;\tau_2,\sigma_2;\tau_3,\sigma_3;\tau_4,\sigma_4)\\
&&\quad\quad=\big\langle{\cal T}\bar{a}_{{\bf l}\sigma_1}(\tau_1)a_{{\bf l}\sigma_2}(\tau_2) \bar{a}_{{\bf l}\sigma_3}(\tau_3)a_{{\bf l}\sigma_4}(\tau_4)\big\rangle_0\\
&&\quad\quad-\big\langle{\cal T}\bar{a}_{{\bf l}\sigma_1}(\tau_1)a_{{\bf l}\sigma_2}(\tau_2)\big\rangle\big\langle{\cal T}\bar{a}_{{\bf l}\sigma_3}(\tau_3)a_{{\bf l}\sigma_4}(\tau_4)\big\rangle_0\\
&&\quad\quad+\big\langle{\cal T}\bar{a}_{{\bf l}\sigma_1}(\tau_1)a_{{\bf l}\sigma_4}(\tau_4)\big\rangle\big\langle{\cal T}\bar{a}_{{\bf l}\sigma_3}(\tau_3)a_{{\bf l}\sigma_2}(\tau_2)\big\rangle_0,
\end{eqnarray*}
where the subscript 0 near brackets indicates that time dependencies and averages are determined by the local operator ${\cal H}_{\bf l}=\sum_\sigma\left[(U/2)n_{\bf l\sigma}n_{\bf l,-\sigma}- \mu n_{\bf l\sigma}\right]$. Due to the translation symmetry the cumulants are identical on all lattice sites, and Fourier transforms of diagrams (a) and (b) are independent of momentum. As was shown in Refs.~\cite{Vladimir,Pairault,Sherman06}, the irreducible part containing these two terms describes the Mott metal-insulator transition. Besides, it was shown \cite{Sherman15} that spectral functions calculated in this approximation are in reasonable agreement with Monte Carlo results for $T\gtrsim t$. Diagrams (c) and (d) are also local and for $U\gg t$ give small corrections to diagrams (a) and (b).

\begin{figure}[t]
\centerline{\resizebox{0.95\columnwidth}{!}{\includegraphics{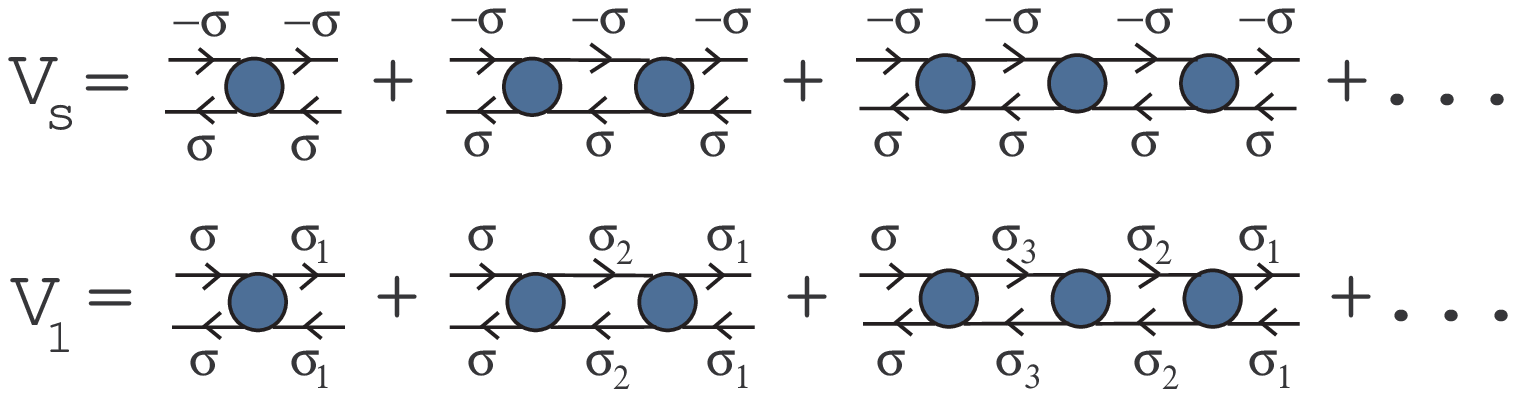}}}
\caption{Two types of infinite sums of ladders contributing to susceptibilities and, as diagram fragments, to $K({\bf k},j)$. In $V_1$, summations over intermediate spin projections are implied.} \label{Fig2}
\end{figure}

In this work, in addition to diagrams (a) and (b) we take into account an infinite sequence of diagrams containing ladder inserts. Several diagrams of this type are shown in the second row in Fig.~\ref{Fig1}. These diagrams are of interest, since sums of ladders define charge and spin susceptibilities \cite{Sherman07},
\begin{eqnarray}\label{susceptibilities}
&&\chi_c({\bf k},\nu)=-\frac{T}{N}\sum_{{\bf q}j}G({\bf q},j)G({\bf k+q},\nu+j)\nonumber\\
&&\quad\quad\quad-\frac{T^2}{N^2}\sum_{{\bf qq'}jj'}\Pi({\bf q},j)\Pi({\bf q'},j')\nonumber\\
&&\quad\quad\quad\times\Pi({\bf k+q},\nu+j)\Pi({\bf k+q'},\nu+j')\nonumber\\
&&\quad\quad\quad\times V_c({\bf q},j;{\bf q'},j';{\bf k+q'},\nu+j';{\bf k+q},\nu+j), \nonumber\\[-0.5ex]
&&\\[-0.5ex]
&&\chi_s({\bf k},\nu)=-\frac{T}{N}\sum_{{\bf q}j}G({\bf q},j)G({\bf k+q},\nu+j)\nonumber\\
&&\quad\quad\quad-\frac{T^2}{N^2}\sum_{{\bf qq'}jj'}\Pi({\bf q},j)\Pi({\bf q'},j')\nonumber\\
&&\quad\quad\quad\times\Pi({\bf k+q},\nu+j)\Pi({\bf k+q'},\nu+j')\nonumber\\
&&\quad\quad\quad\times V_s({\bf q},j;{\bf q'},j';{\bf k+q'},\nu+j';{\bf k+q},\nu+j), \nonumber
\end{eqnarray}
where $\Pi({\bf q},j)=1+t_{\bf k}G({\bf q},j)$ is the terminal line, $N$ is the number of sites, $V_c$ and $V_s$ are the sums of ladders of the type shown in Fig.~\ref{Fig2}.
In the general case circles in Fig.~\ref{Fig2} denote the sum of all four-leg diagrams, which cannot be divided into two disconnected parts by cutting two horizontal particle-hole hopping lines $V_{\rm ir}^{ph}$. The quantities $V_c$ and $V_1$ are connected by the relation $V_c=\sum_{\sigma_1}V_1(\sigma,\sigma_1)$. $V_c$ and $V_s$ are independent of the spin projection for the considered Hamiltonian. Hence the diagrams with ladder inserts describe interactions of electrons with spin and charge fluctuations.

In this work the above sum of irreducible four-leg diagrams $V_{\rm ir}^{ph}$ is approximated by its lowest-order term -- the second-order cumulant $C_2$. As follows from results of Ref.~\cite{Sherman07}, it is a reasonable approximation -- static susceptibilities and some other quantities calculated with its help for small lattices were in satisfactory agreement with Monte Carlo data. As a result the irreducible part reads
\begin{eqnarray}\label{K}
&&K({\bf k},j)=C_1(j)\nonumber\\
&&\quad\quad-\frac{T}{N}\sum_{{\bf k'}j'}\theta({\bf k'},j')\big[V_{s,\bf k- k'}(j,\sigma;j,\sigma;j',-\sigma;j',-\sigma)\nonumber\\
&&\quad\quad+V_{1,\bf k-k'}(j,\sigma;j,\sigma;j',\sigma;j',\sigma)\big]\nonumber\\
&&\quad\quad+\frac{T^2}{2N^2}\sum_{{\bf k'}j'\nu}\theta({\bf k'},j'){\cal T}_{\bf k-k'}(j+\nu,j'+\nu) \nonumber\\
&&\quad\quad\times\Big[C_2(j,\sigma;j+\nu,\sigma;j'+\nu,-\sigma;j',-\sigma)\nonumber\\
&&\quad\quad\times C_2(j+\nu,\sigma;j,\sigma;j',-\sigma;j'+\nu,-\sigma)\nonumber\\
&&\quad\quad+\sum_{\sigma'} C_2(j,\sigma;j+\nu,\sigma';j'+\nu,\sigma';j',\sigma)\nonumber\\
&&\quad\quad\times C_2(j+\nu,\sigma';j,\sigma;j',\sigma;j'+\nu,\sigma')\Big],
\end{eqnarray}
where sums of ladder diagrams $V_s$ and $V_1$, Fig.~\ref{Fig2}, satisfy the Bethe-Salpeter equations (BSE)
\begin{eqnarray}
&&V_{s{\bf k}}(j+\nu,\sigma;j,\sigma;j',-\sigma;j'+\nu,-\sigma)\nonumber\\
&&\;=C_2(j+\nu,\sigma;j,\sigma;j',-\sigma;j'+\nu,-\sigma)\nonumber\\
&&\;+T\sum_{\nu'}C_2(j+\nu,\sigma;j+\nu',\sigma;j'+\nu',-\sigma;j'+\nu,-\sigma)\nonumber\\
&&\;\times{\cal T}_{\bf k}(j+\nu',j'+\nu')\nonumber\\
&&\;\times V_{s{\bf k}}(j+\nu',\sigma;j,\sigma;j',-\sigma;j'+\nu',-\sigma), \label{Vs}\\
&&V_{1{\bf k}}(j+\nu,\sigma';j,\sigma;j',\sigma;j'+\nu,\sigma')\nonumber\\
&&\;=C_2(j+\nu,\sigma';j,\sigma;j',\sigma;j'+\nu,\sigma')\nonumber\\
&&\;+T\sum_{\nu'\sigma''}C_2(j+\nu,\sigma';j+\nu',\sigma'';j'+\nu',\sigma''; j'+\nu,\sigma')\nonumber\\
&&\;\times{\cal T}_{\bf k}(j+\nu',j'+\nu')\nonumber\\
&&\;\times V_{1{\bf k}}(j+\nu',\sigma'';j,\sigma;j',\sigma;j'+\nu',\sigma''), \label{V1}
\end{eqnarray}
and ${\cal T}_{\bf k}(j,j')=N^{-1}\sum_{\bf k'}\theta({\bf k+k'},j)\theta({\bf k'},j')$. The last sum in the right-hand side of Eq.~(\ref{K}) takes into account the multiplier $1/2$ in the diagram (e) in Fig.~\ref{Fig1}. In the used approximation the sums $V_s$ and $V_1$, apart from frequencies, depend only on the transfer momentum.

As  indicated above, the vertex $V_{c\bf k}$ coincides with the symmetrized part of $V_{1\bf k}$. Its antisymmetrized part $\sigma\sum_{\sigma'}\sigma' V_{1\bf k}(\sigma',\sigma,\sigma,\sigma')$ can be shown to coincide with $V_{s\bf k}$.

Equations describing second-order cumulants are given in Refs.~\cite{Vladimir,Pairault,Sherman06,Sherman07}. They are rather cum\-ber\-so\-me. However, the equations can be significantly simplified in the case
\begin{equation}\label{condition}
T\ll\mu,\quad T\ll U-\mu.
\end{equation}
For $U\gg T$ this range of $\mu$ contains the most interesting cases of half-filling, $\mu=U/2$, and moderate doping. For the conditions (\ref{condition}) the first- and second-order cumulants read
\begin{eqnarray}
&&C_1(j)=\frac{1}{2}\left[g_1(j)+g_2(j)\right],\nonumber\\
&&C_2(j+\nu,\sigma;j,\sigma';j',\sigma';j'+\nu,\sigma)\nonumber\\[-1.5ex]
&&\label{cumulants}\\[-1.5ex]
&&\quad=\frac{1}{4T}\big[\delta_{jj'} \big(1-2 \delta_{\sigma\sigma'}\big)
+\delta_{\nu 0}\big(2-\delta_{\sigma\sigma'}\big)\big]\nonumber\\
&&\quad\times a_1(j'+\nu)a_1(j)-\delta_{\sigma,-\sigma'}B(j,j',\nu),\nonumber
\end{eqnarray}
where
\begin{eqnarray*}
&&g_1(j)=(i\omega_j+\mu)^{-1},\quad g_2(j)=(i\omega_j+\mu-U)^{-1}, \\
&&B(j,j',\nu)=\frac{1}{2}\big[a_1(j'+\nu)a_2(j,j') \\
&&\;+a_2(j'+\nu,j+\nu)a_1(j)+a_4(j'+\nu,j+\nu)a_3(j,j')\\
&&\;+a_3(j'+\nu,j+\nu)a_4(j,j')\big], \\
&&a_1(j)=g_1(j)-g_2(j),\quad a_2(j,j')=g_1(j)g_1(j'),\\
&&a_3(j,j')=g_2(j)-g_1(j'),\quad a_4(j,j')=a_1(j)g_2(j').
\end{eqnarray*}

Substituting (\ref{cumulants}) into (\ref{Vs}) we get
\begin{eqnarray}\label{Vsnew}
&&V_{s{\bf k}}(j+\nu,j,j',j'+\nu)=\frac{1}{2}f_1({\bf k},j+\nu,j'+\nu)\nonumber\\
&&\quad\times\bigg\{2C_2(j+\nu,\sigma;j,\sigma;j',-\sigma;j'+\nu,-\sigma)\nonumber\\
&&\quad+\bigg[a_2(j'+\nu,j+\nu)-\frac{\delta_{jj'}}{T}a_1(j'+\nu)\bigg]y_1({\bf k},j,j') \nonumber\\
&&\quad+a_1(j'+\nu)y_2({\bf k},j,j')+a_4(j'+\nu,j+\nu)y_3({\bf k},j,j')\nonumber\\
&&\quad+a_3(j'+\nu,j+\nu)y_4({\bf k},j,j')\bigg\},
\end{eqnarray}
where
\begin{eqnarray*}
&&f_1({\bf k},j,j')=\bigg[1+\frac{1}{4}a_1(j)a_1(j'){\cal T}_{\bf k}(j,j')\bigg]^{-1},\\
&&y_i({\bf k},j,j')=T\sum_\nu a_i(j+\nu,j'+\nu){\cal T}_{\bf k}(j+\nu,j'+\nu)\\
&&\quad\times V_{s{\bf k}}(j+\nu,j,j',j'+\nu).
\end{eqnarray*}
Equations for $y_i({\bf k},j,j')$ follow from Eq.~(\ref{Vsnew}),
\begin{eqnarray}\label{eq_for_y}
&&y_i({\bf k},j,j')=b_i({\bf k},j,j')+\bigg[c_{i2}({\bf k},j,j')\nonumber\\
&&\;-\frac{\delta_{jj'}}{T}c_{i1}({\bf k},j,j')\bigg] y_1({\bf k},j,j')+c_{i1}({\bf k},j,j')y_2({\bf k},j,j')\nonumber\\
&&\;+c_{i4}({\bf k},j,j')y_3({\bf k},j,j')+c_{i3}({\bf k},j,j')y_4({\bf k},j,j'),
\end{eqnarray}
where
\begin{eqnarray*}
&&b_i({\bf k},j,j')=-\frac{1}{4}a_i(j,j')a_1(j)a_1(j'){\cal T}_{\bf k}(j,j')f_1({\bf k},j,j')\\
&&\;+\bigg[a_2(j,j')-\frac{\delta_{jj'}}{T}a_1(j)\bigg]c_{i1}({\bf k},j,j')\\
&&\;+a_1(j)c_{i2}({\bf k},j,j')+a_4(j,j')c_{i3}({\bf k},j,j')\\
&&\;+a_3(j,j')c_{i4}({\bf k},j,j'),\\
&&c_{ii'}({\bf k},j,j')=\frac{T}{2}\sum_\nu a_i(j+\nu,j'+\nu)a_{i'}(j'+\nu,j+\nu)\\
&&\quad\times{\cal T}_{\bf k}(j+\nu,j'+\nu)f_1({\bf k},j+\nu,j'+\nu).
\end{eqnarray*}
Thus, the solution of the BSE (\ref{Vs}) was reduced to the solution of the system of four linear equations (\ref{eq_for_y}) with respect to four variables $y_i({\bf k},j,j')$. The equations depend parametrically on $j$, $j'$ and ${\bf k}$.

In the same manner the BSE for the vertex $V_c$ can be solved. Again using Eq.~(\ref{cumulants}) it can be rewritten as
\begin{eqnarray}\label{Vcnew}
&&V_{c\bf k}(j+\nu,j,j',j'+\nu)=\frac{1}{2}f_2({\bf k},j+\nu,j'+\nu)\nonumber\\
&&\quad\times\Big[2\sum_{\sigma'}C_2(j+\nu,\sigma';j,\sigma;j',\sigma;j'+\nu,\sigma')\nonumber\\
&&\quad-a_2(j'+\nu,j+\nu)z_1({\bf k},j,j')-a_1(j'+\nu)z_2({\bf k},j,j')\nonumber\\
&&\quad-a_4(j'+\nu,j+\nu)z_3({\bf k},j,j')\nonumber\\
&&\quad-a_3(j'+\nu,j+\nu)z_4({\bf k},j,j')\Big],
\end{eqnarray}
where
\begin{eqnarray*}
&&f_2({\bf k},j,j')=\bigg[1-\frac{3}{4}a_1(j)a_1(j'){\cal T}_{\bf k}(j,j')\bigg]^{-1},\\
&&z_i({\bf k},j,j')=T\sum_\nu a_i(j+\nu,j'+\nu){\cal T}_{\bf k}(j+\nu,j'+\nu)\\
&&\quad\times V_{c\bf k}(j+\nu,j,j',j'+\nu),
\end{eqnarray*}
and four quantities $z_i({\bf k},j,j')$ satisfy the system of four linear equations
\begin{eqnarray}\label{eq_for_z}
&&z_i({\bf k},j,j')=d_i({\bf k},j,j')-e_{i2}({\bf k},j,j')z_1({\bf k},j,j')\nonumber\\
&&\quad-e_{i1}({\bf k},j,j')z_2({\bf k},j,j')-e_{i4}({\bf k},j,j')z_3({\bf k},j,j')\nonumber\\
&&\quad-e_{i3}({\bf k},j,j')z_4({\bf k},j,j')
\end{eqnarray}
with
\begin{eqnarray*}
&&d_i({\bf k},j,j')=\frac{3}{4}a_i(j,j')a_1(j)a_1(j'){\cal T}_{\bf k}(j,j')f_2({\bf k},j,j')\\
&&\quad-a_2(j,j')e_{i1}({\bf k},j,j')-a_1(j)e_{i2}({\bf k},j,j')\\
&&\quad-a_4(j,j')e_{i3}({\bf k},j,j')-a_3(j,j')e_{i4}({\bf k},j,j'),\\
&&e_{ii'}({\bf k},j,j')=\frac{T}{2}\sum_\nu a_i(j+\nu,j'+\nu)a_{i'}(j'+\nu,j+\nu)\\
&&\quad\times{\cal T}_{\bf k}(j+\nu,j'+\nu)f_2({\bf k},j+\nu,j'+\nu).
\end{eqnarray*}

Equations (\ref{Larkin}), (\ref{K}), (\ref{Vsnew})--(\ref{eq_for_z}) form a closed set of equations for calculating Green's function (\ref{Green}), which can be solved by iteration. As the starting function in this procedure the result of the Hubbard-I approximation \cite{Hubbard} was used. This function is obtained from the above formulae if $K({\bf k},j)$ is approximated by $C_1$ -- the first term in the right-hand side of (\ref{K}) \cite{Vladimir}. No artificial broadening was used in these calculations.

The maximum entropy method \cite{Press,Jarrell,Habershon} was used for the analytic continuation of calculated Green's functions from the imaginary to real axis.

\section{Results and discussion}
\subsection{The transition to long-range antiferromagnetic order}
\begin{figure}[t]
\centerline{\resizebox{0.9\columnwidth}{!}{\includegraphics{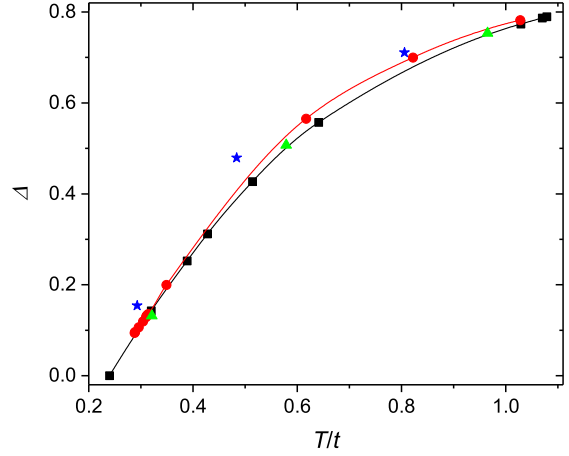}}}
\caption{The temperature dependence of the determinant of the system of linear equations (\protect\ref{eq_for_y}), calculated at half-filling for ${\bf k}=(\pi,\pi)$ and $j=j'$. The black squares and line correspond to $U=8t$, the red circles and line are for $U=5.1t$, the green triangles show results for $U=6t$, while the blue stars for $U=4t$.} \label{Fig3}
\end{figure}

At half-filling, determinants of the systems of linear equations (\ref{eq_for_y}) and (\ref{eq_for_z}) are real. The former determinant $\Delta$ attains its minimum value at ${\bf k}=(\pi,\pi)$ (the intersite distance is set as the length unit) and $j=j'$. With decreasing temperature this value goes down for all considered Hubbard repulsions, as seen in Fig.~\ref{Fig3}.
As follows from the figure, $\Delta$ tends to zero at $T_{\rm AF}\approx 0.24t$, and this value depends only weakly on $U$ and on the lattice size (calculations were carried out in 8$\times$8 and 16$\times$16 lattices). The vanishing determinant leads to the divergence of quantities $y_i$, which entails the divergence of the ladder sum $V_{s\bf k}$ (\ref{Vsnew}) and spin susceptibility $\chi_s({\bf k},\nu)$ (\ref{susceptibilities}) at the antiferromagnetic ordering vector $(\pi,\pi)$ and at the frequency $\omega_{j-j'}=2(j-j')\pi T=0$. Hence the vanishing determinant signals the transition to the long-range antiferromagnetic order.

The finite value of $T_{\rm AF}$ is in contradiction with the Mermin-Wagner theorem \cite{Mermin}. Analyzing equations of the previous section we concluded that this defect is connected with the term proportional to $1/T$ in $C_2$ in Eq.~(\ref{cumulants}). Indeed, this term diverges for $T\rightarrow 0$ indicating that the obtained equations may be inapplicable to very low temperatures. It is the known shortcoming of such series expansions \cite{Oitmaa}. The mentioned weak dependence of $T_{\rm AF}$ on $U$ suggests that unaccounted terms of the series for $K$ have to substitute the factor $1/T$ in the above equations with $1/(T+T_0)$, and the parameter $T_0$ has to be close to $0.24t$. Among these unaccounted terms are diagrams with ladder inserts, in which the role of the longitudinally irreducible vertex $V^{ph}_{\rm ir}$ is played by an infinite sum of transversal ladders constructed from $C_2$ instead of the sole second-order cumulant. Diagrams of this type enabled to obtain close to zero values of $T_{\rm AF}$ in D$\Gamma$A \cite{Schafer} and DF \cite{Rubtsov} approaches. They can be also included into consideration in the SCDT. However, there is also another way, in which the only change in the above equations is the replacement of $1/T$ with $1/(T+T_0)$, where $T_0$ is considered as a correction parameter with a value determined from the condition $T_{\rm AF}=0$. An implicitly similar procedure is used in the D$\Gamma$A approximation \cite{Rohringer} to fulfil this condition. We shall prove this correction in the future work.

The determinant of the second system of linear equations (\ref{eq_for_z}) decreases also with temperature. However, this decrease is much smaller than that in the system (\ref{eq_for_y}), and the determinant never goes to zero. Therefore, the ladder sum $V_{c\bf k}$ (\ref{Vcnew}) and charge susceptibility $\chi_c({\bf k},\nu)$ (\ref{susceptibilities}) do not diverge. This result indicates that there is no charge ordering in the normal-state $t$-$U$ Hubbard model. A deviation from half-filling, which will be considered below, does not change this conclusion.

\subsection{Comparison with results of other methods and sum rules}
\begin{figure}[t]
\centerline{\resizebox{0.9\columnwidth}{!}{\includegraphics{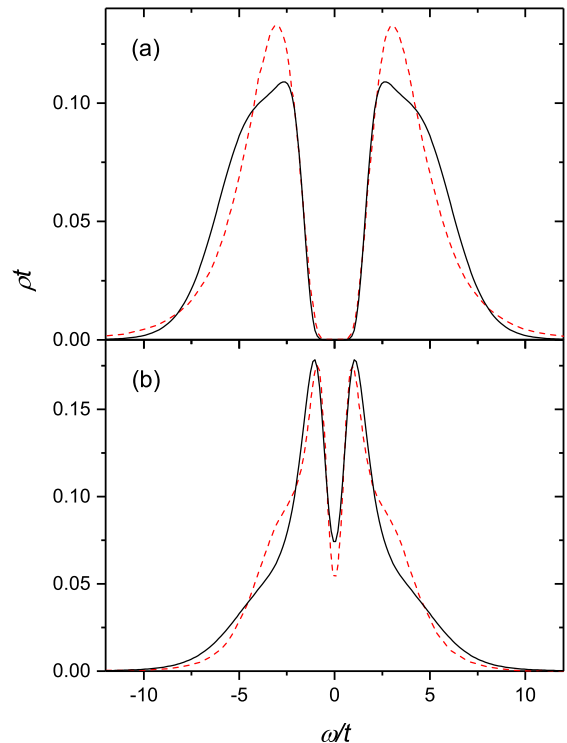}}}
\caption{Densities of states calculated for half-filling, $U=8t$, $T=0.32t$ (a) and $U=4t$, $T=0.29t$ (b) (black solid lines). For comparison results obtained in Monte Carlo simulations \protect\cite{Rubtsov} for $T=0.2t$ and the same repulsions are shown by red dashed lines.} \label{Fig4}
\end{figure}

As mentioned above, spectral functions $A({\bf k},\omega)=-\pi^{-1}{\rm Im}\,G({\bf k},\omega)$ obtained using the SCDT in the lower order approximation \cite{Sherman15} are in satisfactory agreement with Monte Carlo results for $T\gtrsim t$ \cite{Grober}. With account of spin and charge fluctuations one can expect that the agreement can be achieved also for lower $T$. Indeed, Fig.~\ref{Fig4} demonstrates that at half-filling the calculated DOS $\rho(\omega)=N^{-1}\sum_{\bf k}A({\bf k},\omega)$ is in satisfactory agreement with the Monte Carlo data \cite{Rubtsov} for $T\approx 0.2t-0.3t$ (since the Monte Carlo results were obtained at $T=0.2t<T_{\rm AF}$, DOSs calculated for the lowest attained temperatures were used for this comparison). Below spectral functions will be given, which are also close to results of Monte Carlo simulations \cite{Grober} carried at comparatively low $T$.

\begin{figure}[tbh]
\centerline{\resizebox{0.9\columnwidth}{!}{\includegraphics{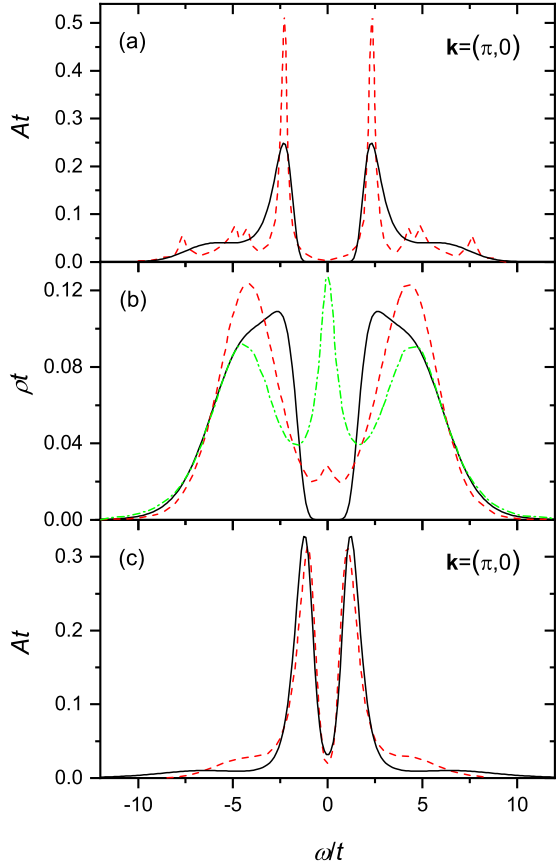}}}
\caption{(a) Spectral functions calculated for half-filling, ${\bf k}=(\pi,0)$ and $U=8t$ by the SCDT (black solid lines, $T\approx 0.32t$) and exact diagonalization \protect\cite{Dagotto} (red dashed line, $T=0$, a 4$\times$4 lattice). (b) Densities of states calculated for half-filling and $U=8t$ by the SCDT (black solid line, $T=0.32t$), the DF approach (red dashed line) and DMFT (green-dash-dotted line). Two latter curves were calculated \protect\cite{Rubtsov} at $T=0.2t$. (c) Spectral functions calculated for half-filling, ${\bf k}=(\pi,0)$ and $U=5.2t$ by the SCDT (black solid lines, $T\approx 0.29t$) and by the DCA with a 64-site cluster \protect\cite{Huscroft} (red dashed line, $T=0.2t$).} \label{Fig5}
\end{figure}
In Fig.~\ref{Fig5} results of the SCDT are compared with some other approaches. As might be expected, the spectral function calculated by exact diagonalization has much more features than our result in panel (a). This can be related to a sparse DOS of the model in a 4$\times$4 cluster used in Ref.~\cite{Dagotto} and to our procedure of analytic continuation, which blurs spectra. Nevertheless, main features of the exact diagonalization spectrum are reproduced correctly by the SCDT. A finite intensity in the Mott gap of the exact-diagonalization spectrum is connected with the artificial broadening $\eta=0.2t$ used in the calculations. Panel (b) demonstrates that the SCDT describes the Mott gap somewhat better than the DF approach \cite{Rubtsov}, in which a finite DOS remains in this range (cf. Fig.~\ref{Fig4}(a)). In the DMFT this parameter set corresponds to a metallic DOS, since this approximation grossly overestimates the critical value of $U$ for the Mott transition. In panel (c), our calculated spectral function is compared with the DCA result \cite{Huscroft}. These spectra are in satisfactory agreement.

\begin{figure}[t]
\centerline{\resizebox{0.9\columnwidth}{!}{\includegraphics{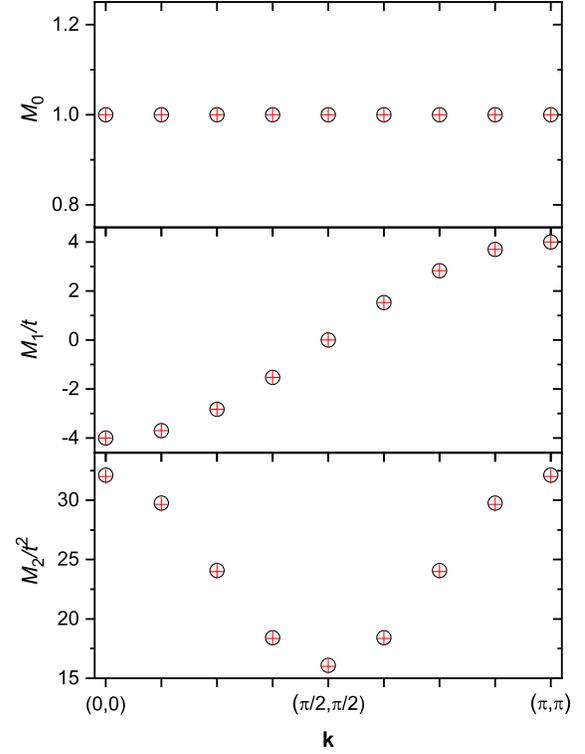}}}
\caption{Moments of the spectral function for half-filling, $U=8t$, $T=0.51t$ and wave vectors on the diagonal of the Brillouin zone (empty circles). Right-hand sides of Eq.~(\protect\ref{moments}) are shown by red crosses.} \label{Fig6}
\end{figure}

As a further test of the used approach the fulfillment of moments sum rules \cite{Kalashnikov,White,Vilk} will be considered. The sum rules read
\begin{eqnarray}
M_0&=&1, \nonumber\\
M_1&=&t_{\bf k}-\mu+\frac{1}{2}U\bar{n}, \label{moments}\\
M_2&=&(t_{\bf k}-\mu)^2+U(t_{\bf k}-\mu)\bar{n}+\frac{1}{2}U^2\bar{n},\nonumber
\end{eqnarray}
where $M_i=\int_{-\infty}^{\infty}\omega^iA({\bf k},\omega)\,{\rm d}\omega$ is the $i$-th moment of the spectral functions and $$\bar{n}=2\langle n_{\bf l\sigma}\rangle=2\int_{-\infty}^{\infty}\rho(\omega)[\exp{(\omega/T)}+1]^{-1}\,{\rm d}\omega$$ is the electron concentration. For the case of half-filling values of the left- and right-hand sides of Eq.~(\ref{moments}) are shown in Fig.~\ref{Fig6}.
As follows from the figure, for this set of parameters the sum rules are fulfilled with a good accuracy.

It should be noted that with decreasing $U$ the difference between left- and right-hand sides of the third equation in (\ref{moments}) grows and runs to 13\% in the zone corner for $U=3t$. This impairment of results is not surprising, since the SCDT was initially devised for strong correlations. However, in this particular case results can be improved if one pays attention that equations (\ref{moments}) have the same structure as the equation
\begin{equation}\label{continuation}
G({\bf k},j)=\int_{-\infty}^{\infty}\frac{A({\bf k},\omega)}{{\rm i}\omega_j-\omega}\,{\rm d}\omega
\end{equation}
used for the analytic continuation. This observation allows one to supplement the above equation with Eq.~(\ref{moments}) and use all them in the extremization procedure of the continuation thus improving the fulfillment of the third sum rule for small repulsions. In the present work this possibility was not used.

\subsection{Four-band structure and phase diagram}
\begin{figure}[t]
\centerline{\resizebox{0.9\columnwidth}{!}{\includegraphics{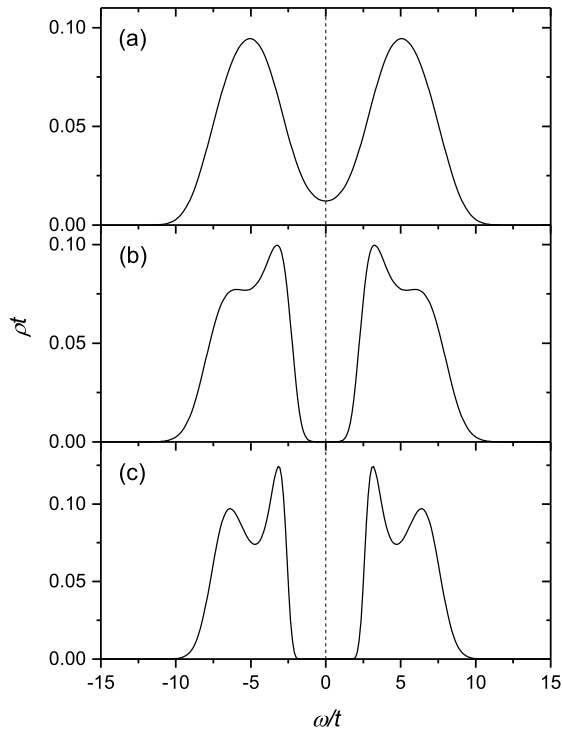}}}
\caption{Densities of states for half-filling, $U=10t$, $T=1.2t$ (a), $0.6t$ (b) and $0.32t$ (c).} \label{Fig7}
\end{figure}

\begin{figure}[hbt]
\centerline{\resizebox{0.9\columnwidth}{!}{\includegraphics{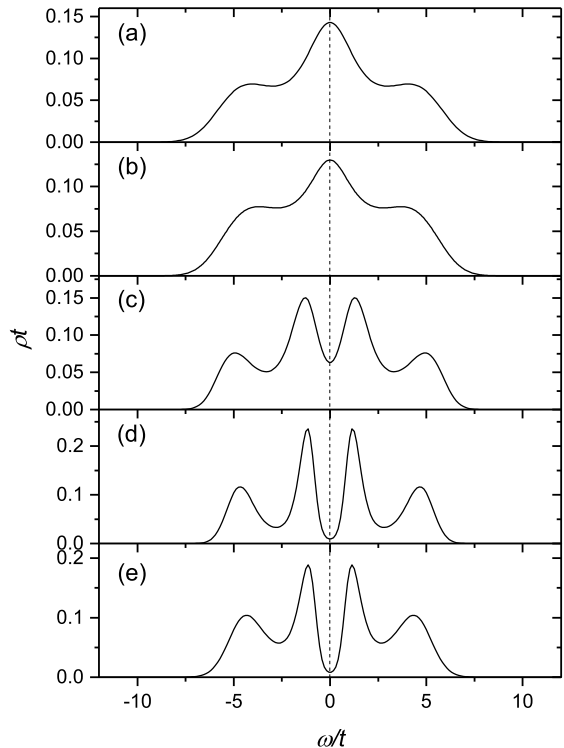}}}
\caption{Same as in Fig.~\protect\ref{Fig7}, but for $U=5.1t$, $T=1.03t$ (a), $0.62t$ (b), $0.49t$ (c), $0.35t$ (d) and $0.29t$ (e).} \label{Fig8}
\end{figure}
\begin{figure}[t]
\centerline{\resizebox{0.9\columnwidth}{!}{\includegraphics{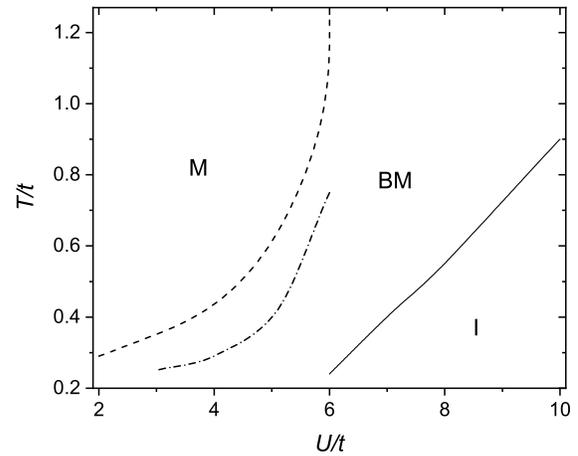}}}
\caption{The location of metallic (M), insulating (I) and bad-metal (BM) states in the $T$-$U$ plane. The dashed line separates metallic and bad-metal states, the solid line segregates the latter and insulating states. The dash-dotted line indicates parameters, for which maxima of internal subbands of the four-band structure fall on the Fermi level.} \label{Fig9}
\end{figure}

Figures \ref{Fig7} and \ref{Fig8} demonstrate temperature dependencies of DOS for strong and moderate repulsions at half-filling. The DOSs in Figs.~\ref{Fig4}(a), \ref{Fig7}(b) and \ref{Fig7}(c) have well-defined Mott gaps around the Fermi level and, therefore, correspond to insulating states. The DOSs in Figs.~\ref{Fig8}(a) and \ref{Fig8}(b) have clear maxima at the Fermi level that relates them to metallic states. Other DOSs have pronounced deeps with finite $\rho$ at the Fermi level. In accord with the used terminology these cases are classified as a bad metal \cite{Emery,Merino,Dasari}. The location of different states in the $T$-$U$ plane is shown in Fig.~\ref{Fig9}.

It is worth noting that the dashed curve in this figure nearly coincides with the metal-insulator boundary found in Ref.~\cite{Sherman17}, in which only diagrams (a), (b) and (e) in Fig.~\ref{Fig1} were taken into account, that is only short-range spin and charge fluctuations were considered. In this approximation, besides critical solutions on the boundary, only two types of states -- metallic and insulating ones -- were found. This result resembles that obtained in DCA \cite{Moukouri}, where also only these two types of states were observed in calculations with 64-site clusters. The shape of the boundary line is similar to that shown in Fig.~\ref{Fig9}. However, in DCA its nearly vertical part is located at much larger repulsions, $U\approx 12t$. Since DCA is a cluster generalization of DMFT, which is known to overestimate the critical repulsion, one can conclude that the former approach does not correct this parameter value. In Fig.~\ref{Fig9}, the location of this nearly vertical part of the boundary, $U\approx 6t$, is close to that obtained in VCA \cite{Balzer}, CDMFT \cite{Park} and second-order DF \cite{Rohringer}. For smaller $U$ the boundary layout resembles that derived in the two-particle self-consistent theory \cite{Vilk}, D$\Gamma$A \cite{Schafer} and the ladder DF \cite{Rohringer}.

From the comparison of Fig.~\ref{Fig9} and Fig.~7 in \cite{Sherman17} it is seen that the long-range fluctuations transform a part of the domain of insulating states into bad-metal states introducing some finite DOS into a Mott gap. Obviously it is connected with strengthening of tails in spectral maxima. Let us consider how this occurs.

\begin{figure}[t]
\centerline{\resizebox{0.9\columnwidth}{!}{\includegraphics{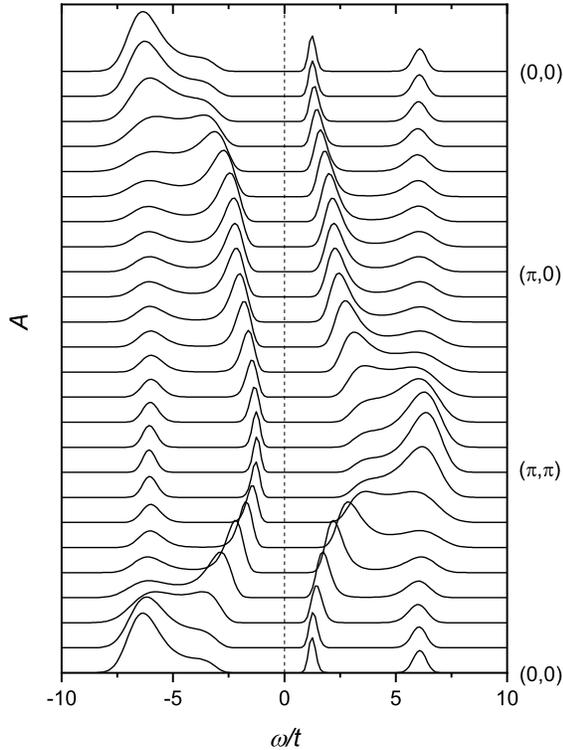}}}
\caption{Spectral functions for momenta located on the symmetry lines of the Brillouin zone. Half-filling, $U=8t$ and $T=0.51t$} \label{Fig10}
\end{figure}
\begin{figure}[t]
\centerline{\resizebox{0.9\columnwidth}{!}{\includegraphics{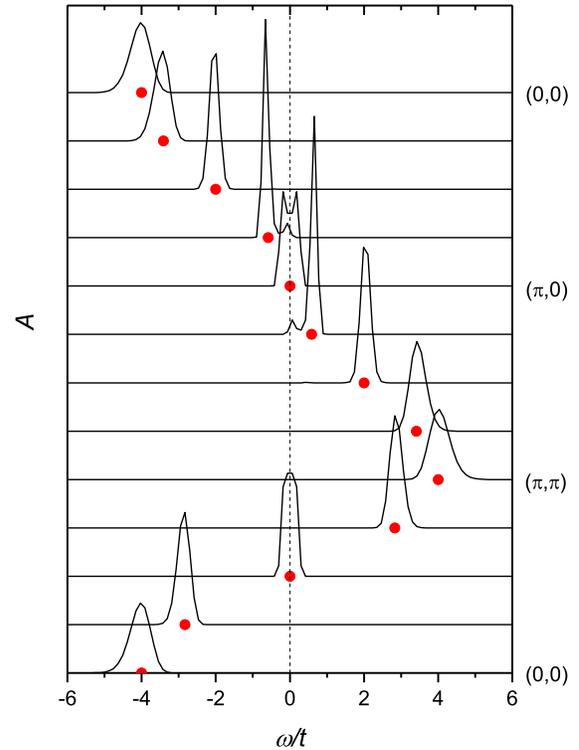}}}
\caption{Same as in Fig.~\protect\ref{Fig10}, but for $U=2t$ and $T=0.24t$. Red dots indicate energies of the uncorrelated electron dispersion $t_{\bf k}=-2t[\cos(k_x)+\cos(k_y)]$ for respective momenta.} \label{Fig11}
\end{figure}

Figure~\ref{Fig10} demonstrates spectral functions along symmetry lines of the Brillouin zone, which are typical for strong repulsions and low temperatures. It should be noticed that shapes of these spectral functions are close to those obtained by Monte Carlo simulations in Ref.~\cite{Grober} for similar parameters (cf. low-temperature spectra in Fig.~1 there). As in the mentioned work, spectra in Fig.~\ref{Fig10} have a pronounced four-band structure. In Refs.~\cite{Sherman15,Sherman16} the respective splitting of the Hubbard subbands and the occurrence of the four-band structure were related to multiple reabsorptions of carriers with the creation of doubly occupied sites. In the following the term ``subband'' will be used in relation to components of the structure. For parameters of Fig.~\ref{Fig10} the Mott gap is well seen between maxima of internal subbands. For a large $U$, the increase of $T$ leads to a broadening of these maxima, which tails fill the gap, as it occurs in Fig.~\ref{Fig7}(a). For a low $T$, with decreasing $U$ maxima of internal subbands are shifted to the Fermi level, and for a small enough $U$ even for low temperatures tails of the maxima fill the Mott gap, as it happened in Figs.~\ref{Fig8}(d) and \ref{Fig8}(e). For further decrease of $U$ internal subbands fall on the Fermi level. It initially occurs at wave vectors $(0,0)$ and $(\pi,\pi)$ for parameters depicted by the dash-dotted curve in Fig.~\ref{Fig9}. This does not lead to the immediate transformation of a dip in the DOS to a maximum at the Fermi level, since spectra for other ${\bf k}$ are at minimum there. For even further decrease of the repulsion a substantial part of internal subbands cross the Fermi level. However, these maxima are kept near the Fermi level and weakened considerably with decreasing $U$, as well as maxima of external subbands. As a result spectral maxima are retained only near frequencies of the uncorrelated spectrum for a major part of momenta. The exception is spectral functions for wave vectors near the boundary of the magnetic Brillouin zone, where both peaks of the former internal subbands are still visible. Now, in the case of weak correlations, they are interpreted as a band splitting due to the antiferromagnetic ordering. Spectral functions in this case are shown in Fig.~\ref{Fig11}.
For half-filling, the boundary of the magnetic Brillouin zone is the one-dimensional (1D) Fermi surface of uncorrelated electrons. The mentioned splitting is seen in Fig.~\ref{Fig11} as dips near the Fermi level for momenta near this surface. Clearly the above discussion describes the transition from the Mott-Hubbard regime of strong electron correlations to the Slater regime of weakly correlated electrons in the presence of long-range antiferromagnetic fluctuations.

For moderate $U$, with increasing $T$ maxima of internal subbands cross the Fermi level, and their intensities are strengthened. This leads finally to the appearance of a maximum of DOS at the Fermi level that is inherent in a metallic state (Figs.~\ref{Fig8}(a) and \ref{Fig8}(b)).

\subsection{Self-energy}
The self-energy can be calculated from the relation
\begin{equation}\label{self-energy}
\Sigma({\bf k},\omega)=\omega-t_{\bf k}-G^{-1}({\bf k},\omega),
\end{equation}
where the imaginary part of Green's function is obtained from the analytic continuation and its real part is derived from this imaginary part and the Kramers-Kronig relation. At half-filling, in insulating states the imaginary part of the self-energy diverges at $\omega=0$ for momenta located on the boundary of the magnetic Brillouin zone. Sharp minima for these wave vectors and frequency are observed also in bad-metal states, especially in cases of a small $\rho(0)$.

\begin{figure}[t]
\centerline{\resizebox{0.9\columnwidth}{!}{\includegraphics{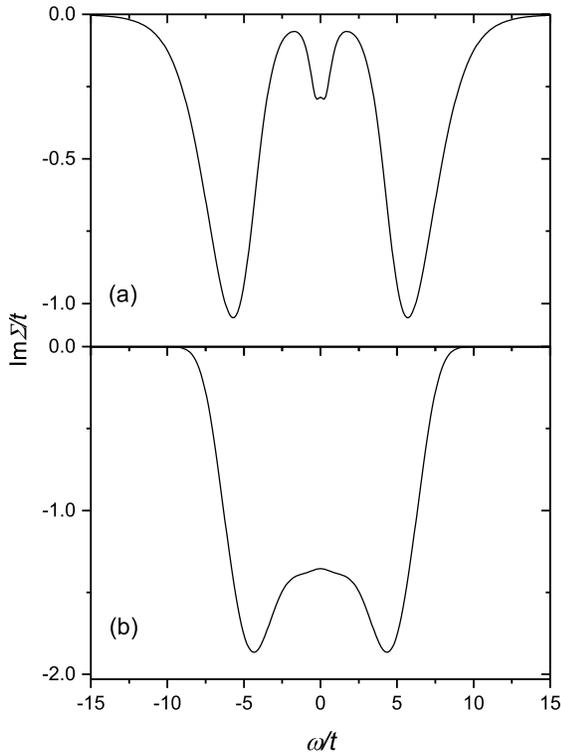}}}
\caption{The imaginary part of the self-energy. Half-filling, ${\bf k}=(\pi,0)$, $U=3t$, $T=0.4t$ (a) and $U=5.1t$, $T=0.62t$ (b).} \label{Fig12}
\end{figure}

Although the SCDT was initially devised for the case of strong correlations, Fig.~\ref{Fig11} demonstrates that the inclusion of spin and charge fluctuations allows one to obtain at least qualitatively correct results also in the case of weak correlations in this approach. Based on this conclusion, let us trace the variation of the self-energy with decreasing $U$ in metallic states at temperatures near their boundary in Fig.~\ref{Fig9}.  Results for two sets of parameters are shown in Fig.~\ref{Fig12} for a momentum on the Fermi surface of uncorrelated electrons (for other wave vectors on this surface ${\rm Im}\,\Sigma({\bf k},\omega)$ looks similar).
The figure illustrates the fact that the electron damping near $\omega=0$ decreases monotonously with reduction in $U$. It can be seen also that in this process for moderate $\omega$ the frequency dependence of the damping tends to $\omega^2$, which is inherent in the Fermi liquid. However, in the considered range of repulsions ${\rm Im}\,\Sigma({\bf k},\omega)$ has a small dip near $\omega=0$ and remains finite in this region.

\subsection{Doping, waterfall, Fermi arc and pseudogap}
\begin{figure}[t]
\centerline{\resizebox{0.9\columnwidth}{!}{\includegraphics{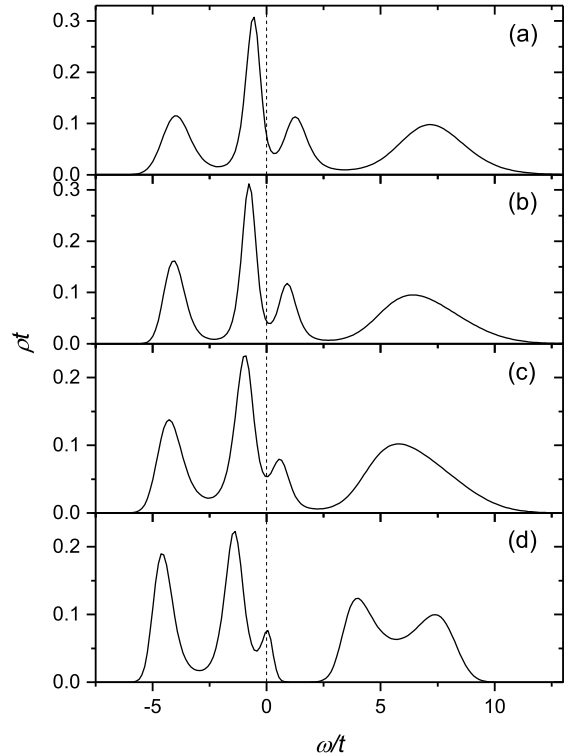}}}
\caption{The density of states for $U=8t$, $T=0.32t$, $\mu=1.5t$ (a, $\bar{n}=0.86$), $\mu=1.7t$ (b, $\bar{n}=0.89$), $\mu=2t$ (c, $\bar{n}=0.93$) and $\mu=2.5t$ (d, $\bar{n}=0.975$).} \label{Fig13}
\end{figure}

Let us consider changes in the electron spectra caused by a deviation from half-filling. Due to the particle-hole symmetry of Hamiltonian (\ref{Hamiltonian}) only the case $\bar{n}<1$ will be investigated. For moderate and strong repulsions doping leads to a strong reconstruction of the DOS. At low temperatures maxima become sharper. For $U\gtrsim 7t$, with doping the intensity is redistributed in favour of the Hubbard subband, in which the Fermi level is located (as an example see Fig.~\ref{Fig13}). An analogous redistribution was observed in Monte Carlo data \cite{Grober}. For moderate $U$ maxima in the DOS of metallic states remain to be pinned to the Fermi level with doping. In this case doping reduces to an intensity decrease below the maxima and an increase above them.

The intensity suppression near $\omega=-2.5t$ in Fig.~\ref{Fig13} originates from one of the pseudogaps, which are responsible for the four-band structure of the DOS near half-filling. This structure arises due to the denominators $\omega+\mu$ and $\omega+\mu-U$ in terms of the SCDT power expansion. As mentioned above, the denominators indicate that the electron scattering and reabsorption are strengthened when the frequency approaches transfer frequencies of the Hubbard atom $-\mu$ and $U-\mu$. As in optics, where an increased reabsorption leads to intensity suppression near respective frequencies, analogous intensity diminutions occur in the considered problem. Hence the pseudogaps are not connected with spin and charge fluctuations – they are observed in spectra of one- \cite{Sherman15} and three-band \cite{Sherman16} Hubbard models derived in approximations without regard for ladder diagrams. It is worth noting that in the latter model one of these pseudogaps plays the role of the Mott gap of the one-band model, separating a Hubbard subband and a band of Zhang-Rice singlets for parameters of hole-doped cuprates.

\begin{figure}[t]
\centerline{\resizebox{0.9\columnwidth}{!}{\includegraphics{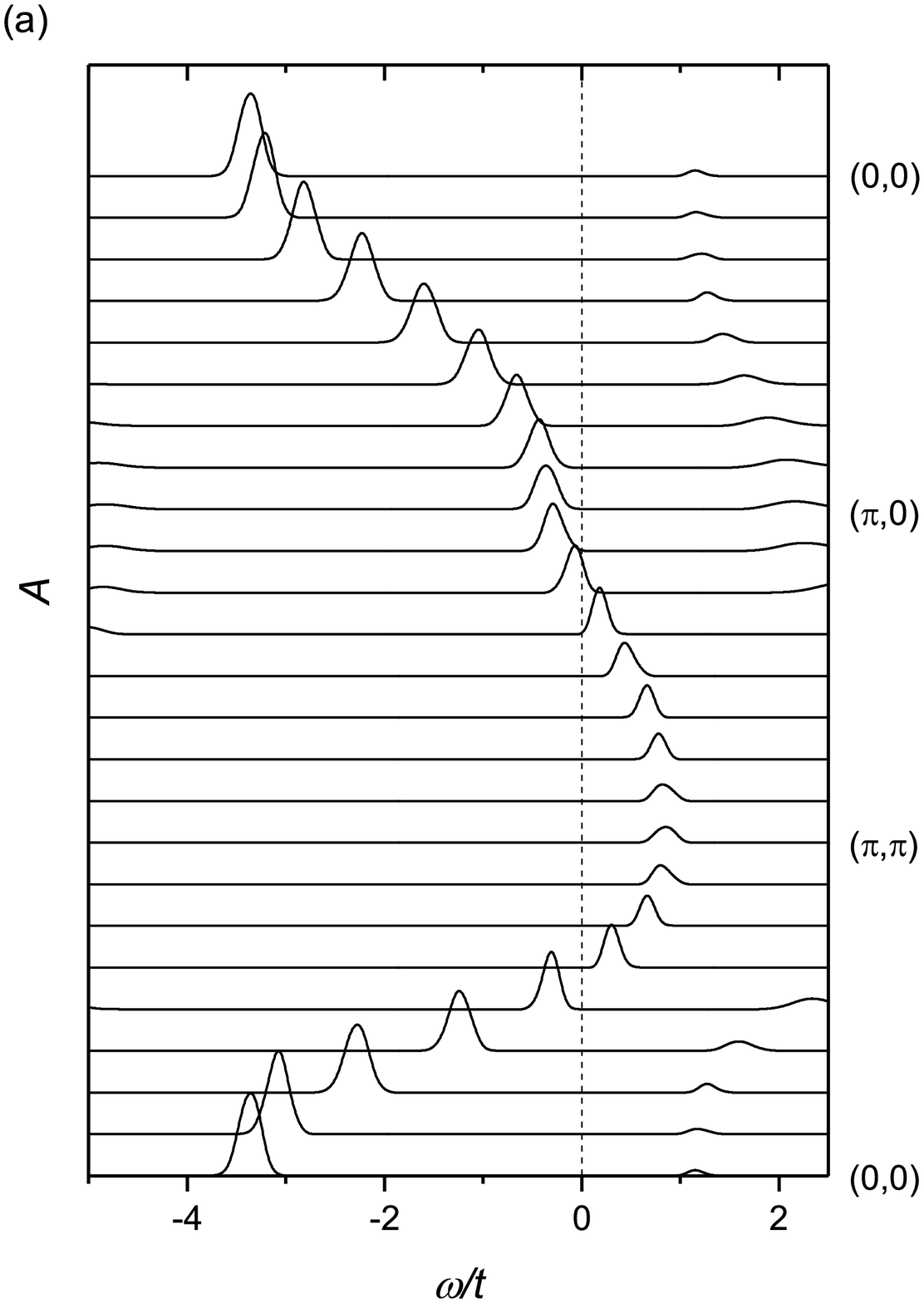}}}
\vspace{1ex}
\centerline{\resizebox{0.9\columnwidth}{!}{\includegraphics{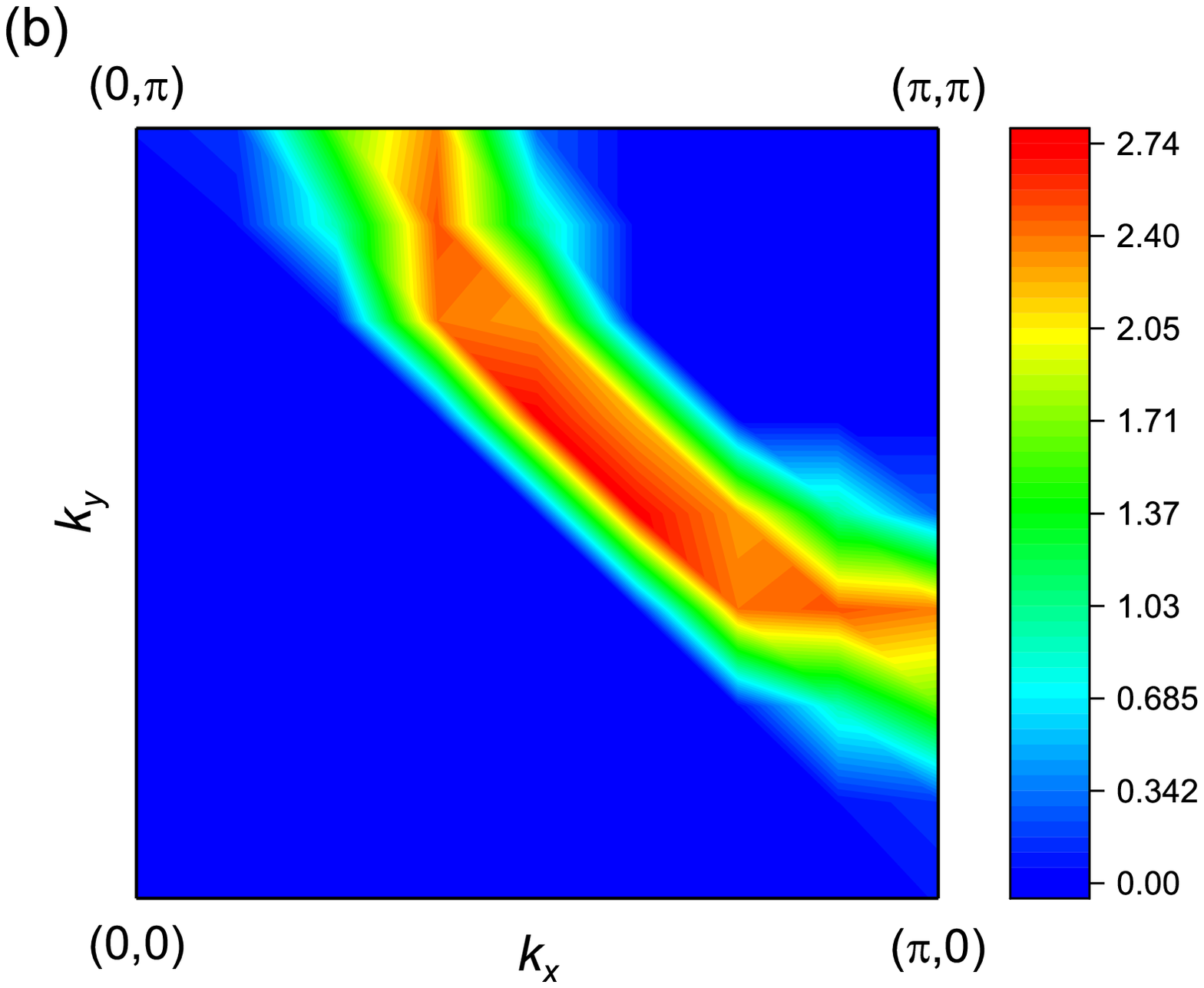}}}
\caption{(a) Spectral functions for momenta located on sym\-met\-ry lines of the Brillouin zone. $U=8t$, $T=0.32t$ and $\mu=1.5t$ ($\bar{n}=0.86$). (b) The contour plot of the spectral function in a narrow frequency window around the Fermi level. $\mu=1.7t$ ($\bar{n}=0.89$), other parameters are the same as in part (a).} \label{Fig14}
\end{figure}
A steep downturn in the electron dispersion, which can be observed in Fig.~\ref{Fig14}(a) in the frequency range from approximately $-3t$ to $-t$, corresponds to the mentioned pseudogap in Fig.~\ref{Fig13}. This dispersion peculiarity is similar to the high-energy anomaly or waterfall observed in photoemission of several families of cuprates \cite{Ronning,Graf,Valla}. As seen from Fig.~\ref{Fig13}, the location of the waterfall depends only weakly on doping, which is in agreement with experimental observations \cite{Graf}. If the exchange constant $J=4t^2/U$ is set to $0.1$~eV, which is close to that observed in hole-doped cuprates, and $U=8t$, the hopping constant $t=0.2$~eV. For parameters of Fig.~\ref{Fig14} the waterfall spans the range from $-0.6$~eV to $-0.2$~eV, which is close to experimental results \cite{Graf}. Thus, the present theory relates the waterfall anomaly to the strong reabsorption of electrons at transfer frequencies of the Hubbard atom. It is worth noting that a similar mechanism for the waterfall formation was suggested in Ref.~\cite{Wang}, in which it was related to the three-site terms of the $t$-$J$ model. These terms describe carrier reabsorption involving intermediate doubly occupied states.

An important consequence of doping is the appearance of a pseudogap near the Fermi level. It is observed for strong repulsions, $U\gtrsim 7t$, and for temperatures close to $T_{\rm AF}$, which points to its intimate relation to antiferromagnetic ordering. As seen in Fig.~\ref{Fig13}, the pseudogap is located near the Fermi level in a wide range of electron concentrations. From this figure it can be also concluded that the pseudogap is a consequence of a splitting out of a narrow band from the internal subband. This mechanism of the pseudogap formation resembles the result obtained in the 2D $t$-$J$ model \cite{Sherman97}, in which the pseudogap was related to the spin-polaron band split out of the Hubbard subband in the antiferromagnetic background. As follows from spectral functions in Fig.~\ref{Fig14}(a), even the dispersion of the segregated band is close to that of the spin-polaron band.

The pseudogap is characterized not only by the dip in the DOS near the Fermi level, but also by the rather specific momentum distribution of intensity. The density plot in Fig.~\ref{Fig14}(b) demonstrates the spectral function integrated in a narrow frequency window around the Fermi level in the underdoped case. In these conditions, the Fermi level crossing occurs along a line segment, which is parallel to the boundary of the magnetic Brillouin zone and is centered at the nodal point located near ${\bf k}=(\pi/2,\pi/2)$. In the considered case the crossing line does not attain boundaries of the Brillouin zone -- as they are approached the intensity fades away. Hence a large rhombus-shaped Fermi surface appears to be disconnected in corners. Similar Fermi surfaces are observed in hole-doped cuprates in the underdoped case \cite{Damascelli}. The line segments forming the non-closed Fermi surface are termed Fermi arcs. As in experiment \cite{Yoshida}, in our calculations their length growth with increasing $1-\bar{n}$, and at the optimal doping $1-\bar{n}\approx 0.15$ arcs attain zone boundaries and the Fermi surface becomes closed. From Fig.~\ref{Fig14}(a) it is seen that the electron dispersion form a plateau near the momentum $(\pi,0)$, which is responsible for the DOS peak at $\omega\approx-0.5t$ in Fig.~\ref{Fig13}(a). Analogous peculiarities are observed in hole-doped cuprates \cite{Damascelli}. Earlier pseudogaps in the spectrum of the Hubbard model were considered using CDMFT \cite{Kyung}, CPT \cite{Senechal04}, DCA \cite{Maier} and VCA \cite{Potthoff}. These methods take into account only short-range spin fluctuations. Therefore, it can be concluded that the pseudogap is mainly connected with these fluctuations.

The model considered above possesses the particle-hole symmetry. In its framework the electron doping produces similar changes in spectra as in the case of hole doping. However, it is well known that spectra of electron-doped cuprates differ essentially from hole-doped brethren \cite{Armitage10}. The point is that a more realistic description of the electron kinetic energy in cuprates needs in introducing the hopping to second and third neighbours that violates the particle-hole symmetry. As was shown in Refs.~\cite{Senechal04,Tohyama}, these terms lead to much stronger antiferromagnetic correlations in electron-doped cuprates in comparison with hole doping. If in the latter case the pseudogap and Fermi arcs remain qualitatively similar to those in the symmetric model, stronger antiferromagnetic correlations in electron-doped cuprates lead to band folding across the boundary of the antiferromagnetic Brillouin zone \cite{Armitage10}. As a consequence there appears a large pseudogap in points corresponding to the intersection of the Fermi surface with this boundary \cite{Armitage10,Armitage01}. Though pseudogaps for hole and electron doping look differently, in both cases they are caused by antiferromagnetic fluctuations.

With doping the determinant of the system of linear equations (\ref{eq_for_y}) remains at a minimum for ${\bf k}=(\pi,\pi)$ and $j=j'$, and the value of $\Delta$ increases with a rise in $1-\bar{n}$. This points to a reduction of the antiferromagnetic correlation length with doping. An analogous behaviour is observed in cuprate perovskites \cite{Birgeneau}. Doping does not perceptibly decrease the de\-ter\-mi\-nant of the system (\ref{eq_for_z}). Hence a deviation from half-filling does not promote a charge ordering.

\section{Concluding remarks}
In this work, the strong coupling diagram technique was used for investigating the influence of spin and charge fluctuations on electron spectra of the 2D $t$-$U$ Hubbard model. The infinite sequences of diagrams with ladder inserts, which were constructed from cumulants of the first and second orders, were taken into account. The sums of these ladders provide a description for spin and charge susceptibilities, and diagrams containing them give an account of electron interactions with spin and charge fluctuations. The obtained equations were solved self-consistently in 8$\times$8 and 16$\times$16 lattices for the ranges of Hubbard repulsions $2t\leq U\leq 10t$, temperatures $0.2t\lesssim T\lesssim t$ and electron concentrations $0.7\lesssim\bar{n}\leq 1$.

At half-filling, the inclusion of spin fluctuations leads to the transition to the long-range antiferromagnetic order at $T_{\rm AF}\approx 0.2t$. In the considered range of $U$ this temperature depends only weakly on the repulsion value and on the lattice size. Calculated densities of states and spectral functions are in satisfactory agreement with results of Monte Carlo simulations, exact diagonalization, DF and DCA calculations. Besides, the spectral functions satisfy moments sum rules with a good accuracy. In the used approach spin and charge fluctuations were considered on the same footing. However, if for the spin subsystem the used approximation leads to antiferromagnetic ordering, no indication of charge ordering was observed in the considered normal-state $t$-$U$ model.

Metallic states are characterized by a pronounced maximum of the DOS at the Fermi level. In the $U$-$T$ plane of Fig.~\ref{Fig9} their existence domain is bounded by the dashed curve. Its part at higher temperatures is close to curves obtained using VCA \cite{Balzer}, CDMFT \cite{Park} and second-order DF \cite{Rohringer}, while the part for lower temperatures resembles the boundary derived in the two-particle self-consistent theory \cite{Vilk}, D$\Gamma$A \cite{Schafer} and ladder DF \cite{Rohringer}. The curve in Fig.~\ref{Fig9} is very close to that obtained in Ref.~\cite{Sherman17}, in which only short-range spin and charge fluctuations were taken into account. In the mentioned work, all states lying outside the domain of metallic states are insulating solutions with well-defined Mott gaps around the Fermi level, as in Ref.~\cite{Moukouri} using DCA with 64-site clusters. In the present work it was shown that the inclusion of long-range fluctuations leads to the appearance of a finite DOS in gaps of a part of insulating solutions lying at smaller $U$ and larger $T$ in Fig.~\ref{Fig9}. Calculated spectral functions show that these finite DOSs appear due to tails of maxima, which are amplified by long-range fluctuations.

Obtained results allowed us to trace the gradual transition from the Mott-Hubbard regime of strong electron correlations to the Slater regime of weakly interacting electrons in the presence of long-range antiferromagnetic correlations. For low temperatures and strong repulsions the electron spectrum has a pronounced four-band structure (see Fig.~\ref{Fig10}), which was earlier observed in Monte Carlo simulations \cite{Preuss,Grober}. In Refs.~\cite{Sherman15,Sherman16} the respective splitting of the Hubbard subbands was related to multiple reabsorptions of carriers with the creation of doubly occupied sites. The transition between the two mentioned regimes proceeds as follows: With decreasing $U$ external maxima of the four-band structure fall off gradually. Internal maxima located near the Fermi level decay also, except for momenta near the boundary of the magnetic Brillouin zone, where both maxima have comparable intensities. For other momenta locations of retained maxima approach frequencies of the uncorrelated electron band. As a result the energy spectrum acquires features of a weakly correlated spectrum except for wave vectors near the boundary of the magnetic Brillouin zone, spectral functions of which have dips near the Fermi level. In metallic states, with decreasing $U$ the damping of electron states falls off gradually at and near the Fermi level. In our approximation, at the Fermi level it remains finite up to the smallest considered values of repulsion.

As follows from obtained results, doping retains the reabsorption pseudogap below the Fermi level. In its properties -- the depressed intensity, the location in the spectrum, the weak dependence of the location on the doping level and the related steep downturn in the electron dispersion -- the pseudogap is similar to high-energy anomalies or waterfalls observed in photoemission of several families of cuprates \cite{Ronning,Graf,Valla}.

Another consequence of the doping, which is seen in the calculated spectra, is the pseudogap near the Fermi level. The pseudogap is observed for repulsions $U\gtrsim 7t$ and for temperatures close to $T_{\rm AF}$ that relates it to the interaction of electrons with antiferromagnetic fluctuations. The appearance of the pseudogap is connected with the splitting out of a narrow band from the Hubbard subband. In this regard the mechanism of the pseudogap formation is close to that in the 2D $t$-$J$ model, in which the pseudogap is also related to the spin-polaron band split out of a Hubbard subband due to the interaction of carriers with antiferromagnetic excitations \cite{Sherman97}. For moderate doping dispersions of the segregated bands are also close. In the underdoped case this dispersion leads to a non-closed Fermi surface consisting of four arcs near $(\pm\pi/2,\pm\pi/2)$. In this surface, intensity fades away as boundaries of the Brillouin zone are approached. With increasing doping the length of the arcs grows, and at the optimal doping $1-\bar{n}\approx 0.15$ the Fermi surface becomes closed and acquires the shape of a rhombus. An analogous evolution of the Fermi surface with doping was observed in photoemission of hole-doped cuprates \cite{Damascelli,Yoshida}.

\ack
I thank the referee who called my attention to the connection between reabsorption pseudogaps and waterfalls. This work was supported by the research project IUT2-27.


\section*{References}
\begin{thebibliography}{99}
\bibitem{Bulut}Bulut N, Scalapino D J and White S R 1993 {\it Phys.\ Rev.} B {\bf 47} 14599
\bibitem{Haas}Haas S, Moreo A and Dagotto E 1995 {\it Phys.\ Rev.\ Lett.} {\bf 74} 4281
\bibitem{Preuss}Preuss R, Hanke W and von der Linden W 1995 {\it Phys.\ Rev.\ Lett.} {\bf 75} 1344
\bibitem{Grober}Gr\"ober C, Eder R and Hanke W 2000 {\it Phys.\ Rev.} B {\bf 62} 4336
\bibitem{Scalapino}Scalapino D 2014 {\it J.\ Phys.: Conference Series} {\bf 529} 012002
\bibitem{Maier}Maier T, Jarrell M, Pruschke T and Hettler M H 2005 {\it Rev.\ Mod.\ Phys.} {\bf 77} 1027
\bibitem{Kyung}Kyung B, Kancharla S S, S\'en\'echal D, Tremblay A-M S, Civelli M and Ko\-t\-liar G 2006 {\it Phys.\ Rev.} B {\bf 73} 165114
\bibitem{Park}Park H, Haule K and Kotliar G 2008 {\it Phys.\ Rev.\ Lett.} {\bf 101} 186403
\bibitem{Sato}Sato T and Tsunetsugu H 2016 {\it Phys.\ Rev.} B {\bf 94} 079907
\bibitem{Senechal00}S\'en\'echal D, Perez D and Pioro-Ladri\`ere M 2000 {\it Phys.\ Rev.\ Lett.} {\bf 84} 522
\bibitem{Senechal04}S\'en\'echal D and Tremblay A-M S 2004 {\it Phys.\ Rev.\ Lett.} {\bf 92} 126401
\bibitem{Kohno}Kohno M 2012 {\it Phys.\ Rev.\ Lett.} {\bf 108} 076401
\bibitem{Huscroft}Huscroft C, Jarrell M, Maier Th, Moukouri S and A. N. Tahvildarzadeh A N 2001 {\it Phys.\ Rev.\ Lett.} {\bf 86} 139
\bibitem{Moukouri}Moukouri S and Jarrell M 2001 {\it Phys.\ Rev.\ Lett.} {\bf 87} 167010
\bibitem{Merino14}Merino J and Gunnarsson O 2014 {\it Phys.\ Rev.} B {\bf 89} 245130
\bibitem{Potthoff}Potthoff M, Aichhorn M and Dahnken C 2003 {\it Phys. Rev.\ Lett.} {\bf 91} 206402
\bibitem{Balzer}Balzer M, Kyung B, S\'en\'echal D, Tremblay A-M S and Potthoff M 2009 {\it Europhys.\ Lett.} {\bf 85} 17002
\bibitem{Arrigoni}Arrigoni E, Aichhorn M, Daghofer M and Hanke W 2009 {\it New J.\ Phys.} {\bf 11} 055066
\bibitem{Faye}Faye J P L and S\'en\'echal D 2017 {\it Phys.\ Rev.} B {\bf 95} 115127
\bibitem{Sherman17}Sherman A 2017 {\it Eur.\ Phys.\ J.} B {\bf 90} 120
\bibitem{Toschi}Toschi A, Katanin A A and Held K 2007 {\it Phys.\ Rev.} B {\bf 75} 045118
\bibitem{Schafer}Sch\"afer T, Geles F, Rost D, Rohringer G, Arrigoni E, Held K, Bl\"umer N, Aichhorn M and Toschi A 2015 {\it Phys.\ Rev.} B {\bf 91} 125109
\bibitem{Rohringer}Rohringer G, Hafermann H, Toschi A, Katanin A A, Antipov A E, Katsnelson M I, Lichtenstein A I, Rubtsov A N and Held K 2017 Diagrammatic routes to non-local correlations beyond dynamical mean field theory arXiv:1705.00024
\bibitem{Rubtsov08}Rubtsov A N, Katsnelson M I and Lichtenstein A I 2008 {\it Phys.\ Rev.} B {\bf 77} 033101
\bibitem{Hafermann}Hafermann H, Li G, Rubtsov A N, Katsnelson M I, Lichtenstein A I and Monien H 2009 {\it Phys.\ Rev.\ Lett.} {\bf 102} 206401
\bibitem{Rubtsov}Rubtsov A N, Katsnelson M I, Lichtenstein A I and Georges A 2009 {\it Phys.\ Rev.} B {\bf 79} 045133.
\bibitem{Georges}Georges A, Kotliar G, Krauth W and Rozenberg M 1996 {\it Rev.\ Mod.\ Phys.} {\bf 68} 13
\bibitem{Hewson}Hewson A C 1993 {\it The Kondo Problem to Heavy Fermions} (Cambridge: Cambridge University Press)
\bibitem{Vladimir}Vladimir M I and Moskalenko V A  1990 {\it Theor.\ Math.\ Phys.} {\bf 82} 301
\bibitem{Metzner}Metzner W 1991 {\it Phys.\ Rev.} B {\bf 43} 8549
\bibitem{Craco}Craco L and Gusm\~{a}o M A 1995 {\it Phys.\ Rev.} B {\bf 52} 17135
\bibitem{Pairault}Pairault S, S\'en\'echal D and Tremblay A-M S 2000 {\it Eur.\ Phys.\ J.} B {\bf 16} 85
\bibitem{Sherman06}Sherman A 2006 {\it Phys.\ Rev.} B {\bf 73} 155105

Sherman A 2015 {\it Physica} B {\bf 456} 35
\bibitem{Sherman15}Sherman A 2015 {\it Int.\ J.\ Mod. Phys.} B {\bf 29} 1550088

Sherman A 2015 {\it Phys.\ Status Solidi} B {\bf 252} 2006
\bibitem{Sherman16}Sherman A 2016 {\it Eur.\ Phys.\ J.} B {\bf 89} 91
\bibitem{Zaitsev}Zaitsev R O 1976 {\it Sov.\ Phys.\ JETP} {\bf 43} 574
\bibitem{Izyumov88}Izyumov Yu A and Skryabin Yu N 1988 {\it Statistical Mechanics of Magnetically Ordered Systems} (New York: Consultants Bureau)
\bibitem{Izyumov90}Izyumov Yu A and Letfulov B M 1990 {\it J.\ Phys.: Cond.\ Mat.} {\bf 1} 8905
\bibitem{Ovchinnikov}Ovchinnikov S G and Valkov V V 2004 {\it Hubbard Operators in the Theory of Strongly Correlated Electrons} (London: Imperial College Press)
\bibitem{Mermin}Mermin N D and Wagner H 1966 {\it Phys.\ Rev.\ Lett.} {\bf 17}, 1133
\bibitem{Dagotto}Dagotto E, Ortolani F and Scalapino D 1992 {\it Phys.\ Rev.} B {\bf 46} 3183
\bibitem{Kalashnikov}Kalashnikov O K and Fradkin E S 1973 {\it Physica Status Solidi} B {\bf 59} 9
\bibitem{White}White S R 1991 {\it Phys.\ Rev.} B {\bf 44} 4670
\bibitem{Vilk}Vilk Y M and Tremblay A-M S 1997 {\it J.\ Phys.\ I France} {\bf 7} 1309
\bibitem{Ronning}Ronning F, Shen K M, Armitage N P, Damascelli A, Lu D H, Shen Z-X, Miller L L and Kim C 2005 {\it Phys.\ Rev.} B {\bf 71} 094518
\bibitem{Graf}Graf J, Gweon G-H, McElroy K, Zhou S Y, Jozwiak C, Rotenberg E, Bill A, Sasagawa T, Eisaki H, Uchida S, Takagi H, Lee D-H and Lanzara A 2007 {\it Phys.\ Rev.\ Lett.} {\bf 98} 067004
\bibitem{Valla}Valla T, Kidd T E, Yin W-G, Gu G D, Johnson P D, Pan Z-H and Fedorov A V 2007 {\it Phys.\ Rev.\ Lett.} {\bf 98} 167003
\bibitem{Sherman97}Sherman A and Schreiber M 1997 {\it Phys.\ Rev.} B {\bf 55} R712
\bibitem{Damascelli}Damascelli A, Hussain Z and Shen Z-X (2003) {\it Rev.\ Mod.\ Phys.} {\bf 75} 473
\bibitem{Yoshida}Yoshida T, Zhou X J, Lu D H, Komiya S, Ando Y, Eisaki H, Kakeshita T, Uchida S, Hussain Z, Shen Z-X and Fujimori A (2007) {\it J.\ Phys.: Condens.\ Matter} {\bf 19} 125209
\bibitem{Hubbard}Hubbard J 1963 {\it Proc.\ R.\ Soc.\ Lond.} A {\bf 276} 238

Hubbard J 1964 {\it Proc.\ R.\ Soc.\ Lond.} A {\bf 277} 237
\bibitem{Sherman07}Sherman A and Schreiber M 2007 {\it Phys.\ Rev.} B {\bf 76} 245112

Sherman A and Schreiber M 2008 {\it Phys.\ Rev.} B {\bf 77} 155117
\bibitem{Press}Press W H, Teukolsky S A, Vetterling W T and Flannery B P 1995 {\it Numerical Recipes in Fortran} (Cambridge: Cambridge University Press) chapter 18
\bibitem{Jarrell}Jarrell M and Gubernatis J E 1996 {\it Phys.\ Rept.} {\bf 269} 133
\bibitem{Habershon}Habershon S, Braams B J and Manolopoulos D E 2007 {\it J.\ Chem.\ Phys.} {\bf 127} 174108
\bibitem{Oitmaa}Oitmaa J, Hamer C and Zheng W 2006 {\it Series Expansion Methods for Strongly Interacting Lattice Models} (Cambridge: Cambridge University Press)
\bibitem{Emery}Emery V J and Kivelson S A 1995 {\it Phys.\ Rev.\ Lett.} {\bf 74} 3253
\bibitem{Merino}Merino J and McKenzie R H 2000 {\it Phys.\ Rev.} B {\bf 61} 7996
\bibitem{Dasari}Dasari N, Vidhyadhiraja N S, Jarrell M and McKenzie R H 2017 {\it Phys.\ Rev.} B {\bf 95} 165105
\bibitem{Slater}Slater J C 1951 {\it Phys.\ Rev.} {\bf 82} 538
\bibitem{Wang}Wang Y, Wohlfeld K, Moritz B, Jia C J, van Veenendaal M, Wu K, Chen C-C and T. P. Devereaux T P (2015) {\it Phys.\ Rev.} B {\bf 92} 07511
\bibitem{Armitage10}Armitage N P, Fournier P and Green R L (2010) {\it Rev.\ Mod.\ Phys.} {\bf 82} 2421
\bibitem{Tohyama}Tohyama T (2004) {\it Phys.\ Rev.} {\bf 70} 174517
\bibitem{Armitage01}Armitage N P,Lu D H, Kim C, Damascelli A, Shen K M, Ronning F, Feng D L, Bogdanov P, Shen Z-X, Onose Y, Taguchi Y, Tokura Y, Mang P K, Kaneko N and Greven M (2001) {\it Phys.\ Rev.\ Lett.} {\bf 87} 147003
\bibitem{Birgeneau}Birgeneau R J and Shirane G 1989 in {\it Physical Properties of High Temperature Superconductors} ed  D M Ginsberg (Singapore: World Scientific)
\end{thebibliography}
\end{document}